\documentclass[]{acta}
\inputencoding{utf8}
\usepackage{supertabular,lscape,epsfig}
\usepackage{amssymb}
\usepackage{amsmath}
\usepackage[T1]{fontenc}
\usepackage{upgreek}
\usepackage{graphicx}
\usepackage{array}
\usepackage{multirow}

\usepackage[backend=biber,style=authoryear,maxbibnames=99,giveninits=true,natbib=true]{biblatex}
\defbibenvironment{references}
    {\begin{list}{}{\leftmargin=0pt \itemsep=1em}}
    {\end{list}}
    {}

\newbibmacro*{refjournal}{%
    \iffieldundef{journaltitle}
        {\iffieldundef{booktitle}
            {\printfield{publisher}}
            {\printfield{booktitle}}}
        {\printfield{journaltitle}}%
}

\renewbibmacro*{begentry}{%
    \refitem
        {\printnames{author}}  
        {\printfield{year}}    
        {\usebibmacro{refjournal}} 
        {\printfield{volume}}  
        {\printfield{pages}}   
}

\DeclareBibliographyDriver{article}{\usebibmacro{begentry}\finentry}
\DeclareBibliographyDriver{book}{\usebibmacro{begentry}\finentry}
\DeclareBibliographyDriver{inbook}{\usebibmacro{begentry}\finentry}
\DeclareBibliographyDriver{incollection}{\usebibmacro{begentry}\finentry}
\DeclareBibliographyDriver{inproceedings}{\usebibmacro{begentry}\finentry}
\DeclareBibliographyDriver{misc}{\usebibmacro{begentry}\finentry}
\SetPages{41}{59}

\SetVol{76}{2026}

\begin{document}

\begin{Titlepage}
\Title{Strange quark star I: the maximum gravitational mass and deformation of magnetized spinning model}
\Author{Fatemeh Kayanikhoo,}
{Nicolaus Copernicus Astronomical Center of the Polish Academy of Sciences, Bartycka 18, 00-716 Warsaw, Poland\\ email: fatima@camk.edu.pl\\
Research Centre for Computational Physics and Data Processing, Institute of Physics, Silesian University in Opava, Bezru\v{c}ovo n\'am.~13, CZ-746\,01 Opava, Czech Republic}

\Author{Mateusz Kapusta,}
{Astronomical Observatory, University of Warsaw, Al. Ujazdowskie 4, 00-478 Warsaw, Poland\\ email: mr.kapusta@student.uw.edu.pl}
\Author{Miljenko \v{C}emelji\'{c},}
{Nicolaus Copernicus Superior School, College of Astronomy and Natural Sciences, Gregorkiewicza 3, 87-100, Toru\'{n}, Poland\\
Nicolaus Copernicus Astronomical Center of the Polish Academy of Sciences, Bartycka 18, 00-716 Warsaw, Poland\\
Research Centre for Computational Physics and Data Processing, Institute of Physics, Silesian University in Opava, Bezru\v{c}ovo n\'am.~13, CZ-746\,01 Opava, Czech Republic}

\Author{Leszek Zdunik,}
{Nicolaus Copernicus Astronomical Center of the Polish Academy of Sciences, Bartycka 18, 00-716
Warsaw, Poland}

\Received{February 2, 2026}

\end{Titlepage}

\Abstract{
We investigate the structural parameters of strange quark stars (SQS) under the influence of strong magnetic fields and varying rotational frequencies. The equation of state is computed using the MIT bag model with a density-dependent bag constant and considering the Landau quantization effect regarding the strong magnetic fields up to $5\times10^{17}\,$G in the interior of SQS.
Employing the LORENE library, we calculate the structural parameters under different magnetic field strengths and rotational frequencies.
Our models are compared in terms of maximum gravitational mass, deformation parameter, binding energy, and compactness. Our equation of state model demonstrates that the gravitational masses are higher than those computed using a MIT bag model with a fixed bag constant. We find the gravitational masses beyond $2.3 \,M_\odot$, which are compatible with the masses of observed compact objects, such as the ``black widow'' pulsar \emph{PSR J0952-0607}, and the \emph{GW190814} event detected by the LIGO/Virgo collaboration. The deformation parameter and maximum gravitational mass of SQS are characterized by fitted functions accounting for variations in both magnetic field strength and rotational frequency. We find the maximum deformation parameter of 1.55 and the maximum gravitational mass of $2.8\, M_\odot$ in the fast-rotating strongly magnetized model. 
}{Pulsars, Compact objects, Relativistic stars, Magnetars}

\section{Introduction}\label{intro}
 Strange quark matter (SQM) is a known stable type of matter which consists of up, down, and strange quarks. The energy per baryon of SQM is comparable to that of $^{56}Fe$, which is approximately 930~MeV \citep{Bodmer_1971, Terazawa_1979, Witten_1984, farhi1984strange,Terazawa_1979a}.

Quarks may become deconfined under extreme conditions of high density ($\geq 10^{15}\, {\rm g/cm^3}$), which can occur in astrophysical environments, for example, in the core of compact objects. Two types of objects containing SQM have been proposed: strange quark stars (SQS) and hybrid stars. The former are hypothetical compact objects made entirely of SQM, while the latter consist of a quark matter core surrounded by a shell of hadronic matter. The study of these exotic types of compact objects is of great interest, as they can provide insights into the properties of matter at extreme densities and temperatures. The stability of SQSs and the stellar parameters of these objects compared to the neutron stars were initially studied by \cite{Alcock_1986} and independently by \cite{Haensel_1986}. \cite{Weber_1997} discussed quark deconfinement in the core of neutron stars and compared the structural parameters of them with neutron stars. 

The scenarios for the phase transition of nuclear matter to quark matter in the cores of compact objects are dependent on specific astrophysical conditions. One possible scenario involves a binary system comprising a neutron star (NS) or a proto-neutron star (proto-NS), wherein the companion star is undergoing Roche lobe overflow and subsequent accretion \citep{Ouyed_2002, Ouyed_2013, Ouyed_2015}. Another possibility is a decrease in the centrifugal force as the NS spins down. In both scenarios, the increase in the central density leads to quark deconfinement and the associated gravitational implosion may be accompanied by a luminous ejection of the envelope, termed a Quark-Nova and the remnant core might be SQS. Conversion of NS to SQS releases the energy in order of $10^{52}\, \mathrm{ergs}$. The Quark-Nova explosions might be sources of the observed X-rays, gamma-rays, and fast radio bursts (FRBs) \citep{Ouyed_2002, Ouyed_2020_GRB, ouyed2022}. 

The recent discovery of a compact object in the supernova remnant \emph{HESS J1731-347}, with the lightest mass observed up to now, may indicate the presence of a composition of exotic matter, possibly quark matter. \cite{Doroshenko_2022} estimated the gravitational mass and radius of this source, using analysis of the X-ray spectrum and distance estimates from Gaia observations, as $M_{\rm g}= 0.77 ^{+0.20}_{-0.17} M_{\odot}$ and $R = 10.40^{+0.86}_{-0.78}\,$km, respectively. However, there is still no direct confirmed detection of quark stars.

The recent detections of compact objects using advanced instruments have provided more accurate information about such objects. One example is the detection of the gravitational waves from merging compact objects \emph{GW190814} by the LIGO-VIRGO collaboration, which carries important information about their interior material and shape. The gravitational mass of the compact is predicted to be about $2.6\,M_\odot$ by \cite{Abbott:2020:}. 

To provide a realistic model, the properties of compact objects should be considered, such as the composition of matter, magnetic field strength, and rotation. Among numbers of EOS models for compact objects, best known for quark stars are the MIT bag model \citep{farhi1984strange,Johnson:1975zp,Hecht2000}, Nambu-Jona-Lasinio (NJL) model \citep{Terazawa_1977, Prov_2009_NJL,Prov_2020_Hyb,Prov_2020_hyb_njl}, and confined density-dependent mass (CDDM) model \citep{CHAKRABARTY1989112,CHAKRABARTY1989112,Hou_2015}, predicting properties of compact objects, such as gravitational mass, radius, and shape. Comparison with the observed parameters can help constrain the EOS, providing information about the nature of matter under extreme conditions.  

According to the theoretical studies the magnetic field in the core of magnetars might reach $10^{18}\,$G \citep{Haensel_1986,Lai_1991,Boc_1995,Isa_2014}. There are a number of studies on the stellar properties of magnetized neutron stars and strange stars including those by \cite{Mallick_2014,Chatterjee_2015,Mastrano_2015} in which the stellar parameters such as maximum gravitational mass, radius, and deformation of NS affected by the strong magnetic field. 

The spin period affects the gravitational mass and radius and causes the deformation of compact objects. \cite{Gondek-Rosinska_2000} studied the rapidly rotating SQS and investigated the effect of rotational frequency on the overall parameters of both neutron stars and quark stars. They showed that, compared to models of neutron stars, rotation has a more significant impact on the overall parameters of strange stars. \cite{Haensel_2009} studied Keplerian frequency of neutron stars and quark stars. They found an approximate formula to estimate the Keplerian frequency for considered stellar models.

In this paper, we study the influence of both magnetic field and rotational frequency on the structural parameters of SQS including gravitational mass, radius, and deformation as well as their total and binding energies. The magnetic field strength ranges $0$ to $5\times 10^{17}\,$G, and the selected rotational frequencies are 0, 400, 800, and 1200~Hz. In the companion paper, we study the Keplerian configuration of SQS. 

The paper is organized as follows: in the Section~\ref{sEOS} is presented the EOS model; in Section~\ref{STR} are described the equations of stellar structure in axisymmetric space-time; in Section~\ref{NR} we introduce the numerical code LORENE and the numerical setups; in Section~\ref{RES} are investigated the computed stellar parameters, as well as the energy of SQS in each model. We summarize and conclude in Section~\ref{CONCL}.
\section{Equation of state}\label{sEOS}
\begin{figure*}
 \centering
  \includegraphics[width=223.857bp]{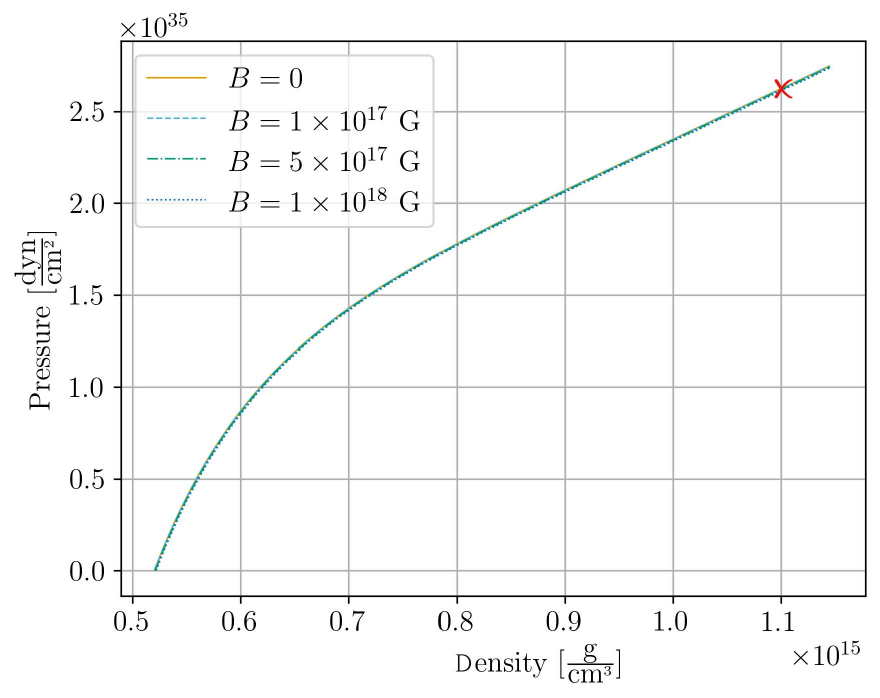}
 \caption{Pressure as a function of the energy density of SQM in the presence of different magnetic field strengths. The energy density corresponding to the maximum gravitational mass is pointed with the red cross. The curves with four magnetic fields indicated in the legend  overlap.}
\label{slEOS}
\end{figure*}
We consider an SQS which contains {\em up}, {\em down} and {\em strange} quarks. The mass of the strange quark is $150$~MeV. The fraction of electrons produced during the phase transition is low, of the order of $10^{-3}$, so we simplify our model by neglecting their contribution. The EOS is computed using the MIT bag model \citep{farhi1984strange,Johnson:1975zp, Hecht2000} with density-dependent bag constant \citep{Burgio_2002}. In this model, the total energy contains the kinetic energy of quarks that is computed from Fermi relations and the bag constant $\mathcal{B}_{\rm bag} (\rho)$,
\begin{equation}
\varepsilon_{\rm tot}=\sum_{i, j=\pm}\varepsilon_{i}^{(j)}+\mathcal{B}_{\rm bag} (\rho),		\label{E03}
\end{equation}
where $j=\pm$ is the spin polarization of quarks (spin up and down, respectively) and $i\in (1,2,3)$ represents {\em up}, {\em down} and {\em strange} quarks.
Due to the strong magnetic field interior of the compact object which in  our study here is up to $5\times10^{17}\,$G, we rewrite the Fermi relations considering the Landau quantization effect \citep{Lan:1977:QM:,Lop:2015:JCAP:,Muk:2017:RRP:}. 
The single particle energy density is  
\begin{equation}
\epsilon^{i}=[p_{i}^{2}c^{2}+m_{i}^{2}c^{4}(1+2 \nu B_{\rm D})]^{1/2},		\label{E01}
\end{equation}
where $p_{i}$ and $m_{i}$ are the momenta and the mass of quark $i$, $\nu$ is the Landau level and $B_{\rm D}=B/B_{\rm C}$ is the dimensionless magnetic field, where $B_{\rm C}=m_{i}^{2}c^{3}/q_{i} \hbar$, with $q_{i}$ the charge of quark $i$. 

The number density of quarks is obtained from
\begin{equation}
\rho=\sum_{\nu=0}^{\nu_{\rm max}} \frac{2qB}{h^{2}c}g(\nu) p_{F}(\nu),		\label{E02}
\end{equation}
where $\nu_{\rm max}$ is the maximum number of Landau levels corresponding to the maximum Fermi energy $\epsilon_{\rm Fmax}$,
\begin{equation}
    \nu_{\rm max}=\frac{\epsilon_{\rm Fmax}^{2}-1}{2 m_{i} c B_{D}} \label{E02_1}.
\end{equation}
In Eq.~\ref{E02}, $g(\nu)$ and $p_{F}(\nu)$ are the degeneracy and Fermi momentum of the $\nu$-th Landau level, respectively.
The kinetic energy density of particles is defined as,
\begin{equation}
\varepsilon_{i}^{(j)}=\frac{2B_{D}}{(2\pi)^{2}\lambda^{3}}m_{i}c^{2}\sum_{\nu=0}^{\nu_{\rm max}} g_{\nu}(1+2\nu B_{D})\eta(x),
\label{E04}
\end{equation}
where
\begin{equation}
\eta(x)=\frac{1}{2}\left[x\sqrt{1+x^{2}}+ln(x\sqrt{1+x^{2}})\right],
\label{E05}
\end{equation}
with
\begin{equation}
x=\frac{X^{(j)}_{F}}{(1+2\nu B_{D})^{1/2}}
\label{E06}
\end{equation}
and 
\begin{equation}
X^{(j)}_{F}=(\epsilon_{F}^{(j)2}-1-2\nu B_{D})^{1/2}.
\label{E07}
\end{equation}

By considering a zero magnetic field, the maximum number of Landau levels $\nu_{\rm max}$ goes to infinity and one can simplify the kinetic energy (Eq.~\ref{E05}) to the Fermi relation.

The density-dependent bag constant $\mathcal{B}_{\rm bag} (\rho)$ is defined with a Gaussian relation 
\begin{equation}
\mathcal{B}_{\mathrm{bag}}(\rho)=\mathcal{B}_{\mathrm{\infty}}+(\mathcal{B}_{0}-\mathcal{B}_{\mathrm{\infty}})e^{-\alpha(\rho/\rho_0)^2},
\label{E08}
\end{equation}
with the number density $\rho_0 = 0.17\, {\rm fm^{-3}}$, $\alpha = 0.17 $ and $\mathcal{B}_{0}=\mathcal{B}_{\rm bag}(0)=400~\mathrm{MeV/fm^{3}}$. The parameter $\mathcal{B}_{\mathrm{\infty}}$ is defined in such a way that the bag constant would be compatible with experimental data of CERN-SPS. $\mathcal{B}_{\mathrm{\infty}} =8.99~\mathrm{MeV/fm^{3}}$ is determined by putting the quark energy density equal to the hadronic energy density \citep[see][and references therein]{LOCV2000, Burgio_2002}.

We can write the relation between pressure $P$ and total energy density $\varepsilon_{\rm tot}$ (EOS) as
\begin{equation}
P(\rho)=\rho \left(\frac{\partial \varepsilon_{\rm tot}}{\partial \rho}\right) - \varepsilon_{\rm tot}.
\label{E09}
\end{equation}
Fig.~\ref{slEOS} shows the pressure $P$ versus the density $\varepsilon_{\rm tot}$ in the presence of different magnetic field strengths. Our model of EOS indicated zero pressure at the central density of $\simeq 0.52 \times 10^{15}$~$\mathrm{gr/cm^3}$ ($\simeq 290\, \mathrm{MeV/fm^3}$) and the maximum central value $1.1 \times 10^{15}$~$\mathrm{gr/cm^3}$ ($\simeq 613\, \mathrm{MeV/fm^3}$) corresponding to the maximum gravitational mass of SQS, which is approximately pointed with a red cross on Fig.~\ref{slEOS}. 

A number of studies have investigated different models of EOS for quark stars. To assess the performance of our model, we compared it to the EOS presented in \cite{Chatterjee_2015}. They provided the linear EOS (MIT bag model with fixed bag constant $\mathcal{B}_{\rm bag} = 60\, \mathrm{MeV/fm^3}$) in which the maximum gravitational mass is inversely proportional to the square root of energy density at zero pressure. We linearize our EOS model and extrapolate the value of energy density to zero pressure, $\approx 2\times 10^{14}\, \mathrm{g/cm^3}$ that is half of the value of the model provided by \cite{Chatterjee_2015}. Consequently, our model predicts a larger maximum gravitational mass, which aligns well with recent observational data. Notably, both our model and Chatterjee's indicate that magnetic fields below $10^{19}\,$G do not significantly impact the stiffness of EOS. However, in the following sections, we show that the structural parameters and shape of the star significantly change with the magnetic field strength. 

\section{The equations of stellar structure}\label{STR}
In this section, we derive the differential equations of the stellar structure by solving the Einstein equation,
\begin{equation}
    R^{\mu \nu} - \frac{1}{2} R g^{\mu \nu} = 8 \pi T^{\mu \nu},
\end{equation}
where $R^{\mu \nu}$, $g^{\mu \nu}$ and $T^{\mu \nu}$ are the Ricci, metric and energy-momentum tensors, respectively, and $R$ is the Ricci scalar. We choose units with $G=c=1$. 

We start with the energy-momentum tensor of the perfect fluid and the spherically symmetric star. Then, regarding the strong magnetic field and the spin of the compact object, we inspect the energy-momentum tensor coupling with the Maxwell energy-momentum tensor and solve the Einstein equation in the axisymmetric space-time.  

The Tolman-Oppenheimer-Volkov equations (TOV) are derived by solving the Einstein field equations in the spherically symmetric space-time for the perfect fluid energy-momentum tensor \citep{Tolman_1939, Open_Vol_1939} 
\begin{equation} \label{perfect}
    T^{\mu \nu} = (\varepsilon+P)u^\mu u^\nu + P g^{\mu \nu},
\end{equation}
where $u^{\mu}$ is the velocity 4-vector, $g^{\mu \nu}$ is metric tensor, $\varepsilon$ is the energy density and $P$ is the pressure of the perfect fluid. The differential equations are as follows, 
\begin{equation} \label{TOV}
    \frac{dP}{dr} = -(P+\varepsilon)\frac{m+4 \pi r^3 P}{r(r-2m)}
\end{equation}
and
\begin{equation}
    \frac{dm}{dr} = 4 \pi r^2 \varepsilon .  
\end{equation}

In the presence of the magnetic field, considering the interaction of the electromagnetic field with the matter (magnetization), we can rewrite the energy-momentum tensor as
\begin{multline}\label{E12}
    T^{\mu \nu}=(\varepsilon + P)u^{\mu}u^{\nu}+P g^{\mu \nu}+\frac {\mathcal{M}}{B}\Big[ b^{\mu}b^{\nu}-(b\cdot b)\\(u^{\mu}u^{\nu}+g^{\mu \nu}) \Big]+
\frac{1}{\mu_{0}}\Big[-b^{\mu}b^{\nu}+(b\cdot b)(u^{\mu}u^{\nu}+\frac{1}{2}g^{\mu \nu})\Big], 	
\end{multline}
where the two first terms are the perfect fluid contribution given by Eq.~\ref{perfect}, the third term is the contribution of the magnetization $\mathcal{M}$ and the last term is the pure magnetic field contribution, where $B$ is the magnetic field and $b^{\mu}$ is the magnetic field 4-vector. The magnetization $\mathcal{M} $ is given by the coupling between the electric current $j^\phi$ and the magnetic vector potential $A_{\phi}$ \citep[for more details see][]{Chatterjee_2015, Franzon_2017}. 

We solve the Einstein field equations within the 3+1 formalism in an axisymmetric space-time, 
\begin{equation}
    ds^{2}=-N^2 dt^2+A^2(dr^2+r^2 d \theta^2)+\lambda^2r^2\sin^2(\theta)(d\phi-N^{\phi}dt)^2		\label{E10}
\end{equation}
where $N$, $A$, $\lambda$, and $N^{\phi}$ are functions of $(r,\theta)$.
By applying 3+1 formalism we obtain a set of four elliptic partial differential equations (PDEs),
\begin{multline}
\Delta_3=4\pi A^2(E^T+S^{r}_{r}+S^{\theta}_{\theta}+S^{\phi}_{\phi})+\frac{\lambda^2 r^2 \sin^2(\theta)}{2N^2}\delta N^{\phi}\delta N^{\phi}\\
-\delta \nu \delta(\nu+\beta)		\label{E11a}
\end{multline}
\begin{equation}
\Delta_2[\alpha+\nu]=8\pi A^2 S^{\phi}_{\phi}+\frac{3\lambda^2 r^2 \sin^2(\theta)}{4N^2}\delta N^{\phi}\delta N^{\phi}-\delta \nu \delta \nu 		\label{E11b}
\end{equation}
\begin{equation}
\Delta_2[(N\lambda -1)r \sin(\theta)]=8\pi N A^2 \lambda r \sin(\theta)(S^{r}_{r}+S^{\theta}_{\theta})		\label{E11c}
\end{equation}
and
\begin{multline}
\left[\Delta_3-\frac{1}{r^2 \sin^2(\theta)}\right](N^{\phi}r \sin(\theta))=-16\pi \frac{NA^2}{\lambda^2}\frac{J^{\phi}}{r \sin(\theta)}+\\
r \sin(\theta)\delta N^{\phi}\delta(\nu-3 \beta),	\label{E11d}
\end{multline}
where $\nu=\ln N$, $\alpha=\ln A$, $\beta=\ln \lambda$, and $J^{\phi}$ is electromagnetic current. In the above equations, $E^T$ and $S^i_j$ are total energy and stress, respectively \citep{Franzon_2017}.

Considering a magnetic field pointing in the z-direction, we can rewrite the energy-momentum tensor Eq.~\ref{E12} in a well-known form as follows,
\begin{multline}
T^{\mu \nu}=\mathrm{diag}\left(\varepsilon +\frac{B^{2}}{2\mu_{0}}
, P-\mathcal{M}B+\frac{B^{2}}{2\mu_{0}},\right.\\
\left. P-\mathcal{M}B +\frac{B^{2}}{2\mu_{0}}, P-\frac{B^{2}}{2\mu_{0}} \right).
\label{E14}
\end{multline}
In Eq.~\ref{E14}, the magnetization term $\mathcal{M}$ reduces the total pressure of the system. It is also clear that the magnetic field reduces the parallel pressure, but the perpendicular pressure increases with increasing the magnetic field. 
\section{The numerical method}\label{NR}
We solve a set of four PDE equations (Eqs.~\ref{E11a}-\ref{E11d}) presented in Section~\ref{STR} by using the LORENE library\footnote{\url{http://www.lorene.obspm.fr}} \citep{Bonazzola_1998, Chatterjee_2015, Franzon_2017}. 

The LORENE library is a code based on C++ that uses the spectral method to solve PDEs. This method is more accurate than grid-based methods, thereby enabling more precise calculations of the solutions to this system. Space in LORENE is separated into domains and mapped onto the specific coordinate system that can be readjusted in order to handle non-spherical shapes. Our setup for the magnetized rotating SQS consists of $3$ domains: one covering the interior of the star, a second forming a thin shell next to the surface, and a third extending outward to infinity. We use \texttt{Et\_magnetisation} class\footnote{The code is located in the directory \texttt{Lorene/Codes/Mag\_eos\_star}.} to calculate hydrostatic configurations for uniformly (not differentially) rotating magnetized stars. 

The magnetic field is specified using the so-called current function, $k_0$, which describes the amplitude of current inside the star to generate the magnetic field \citep{Franzon_2017}. In our setup, the current function amplitude varies from $0$ to $16000$ in the intervals of $2000$, enabling us to cover vast ranges of central magnetic field values up to $5\times 10^{17}\,$G. 

As mentioned earlier, we solve the equations for different rotational frequencies 0, 400, 800, and 1200~Hz. 

In every series of calculations for a given frequency and magnetic field, we compute the parameters of 51 stellar configurations with specified central enthalpy values in the range from $0.01\,c^2$ to $0.51\,c^2$ with the spacing of $0.01\,c^2$ where $c$ is the speed of light. 

To expedite the calculations, we utilized a wrapper based on MPI that facilitates the distribution of the computational load across the available threads. 
By such parallelizing, we significantly enhanced the speed and efficiency of the calculations, resulting in faster processing times and improved performance.

Equilibrium configurations in Newtonian gravity are known to satisfy the virial relation when a polytropic equation of state is assumed. This relation is commonly utilized to verify the accuracy of computations. The 3-dimensional virial identity (GRV3), introduced by \cite{GRV3}, extends the Newtonian virial identity to general relativity. On the other hand, the 2-dimensional virial identity (GRV2) proposed by \cite{GRV2}, generalizes the virial identity for axisymmetric space-times to general asymptotically flat space-times. Our computational results indicate a high level of accuracy which is $\approx 10^{-5}$ in the non-magnetized--non-rotating models and $\approx 10^{-2}$ in the magnetized--fast-rotating model.
\section{Analysis of stellar parameters}\label{RES}
In this section, we examine the properties of the star under varying conditions, where  both the magnetic field strength and rotational frequency are varied. In the following four-panel figures, each panel shows a model with a rotational frequency $f = [0, 400, 800, 1200]\,$Hz and colors show the strength of the central magnetic field which varies from $0$ to $5\times 10^{17}\,$G. Each circle is one computed configuration specified with a central enthalpy. 
\subsection{Gravitational mass and radius}\label{MgR}
\begin{figure*}
\centering
\makebox[\linewidth][c]{\includegraphics[width=368.500bp]{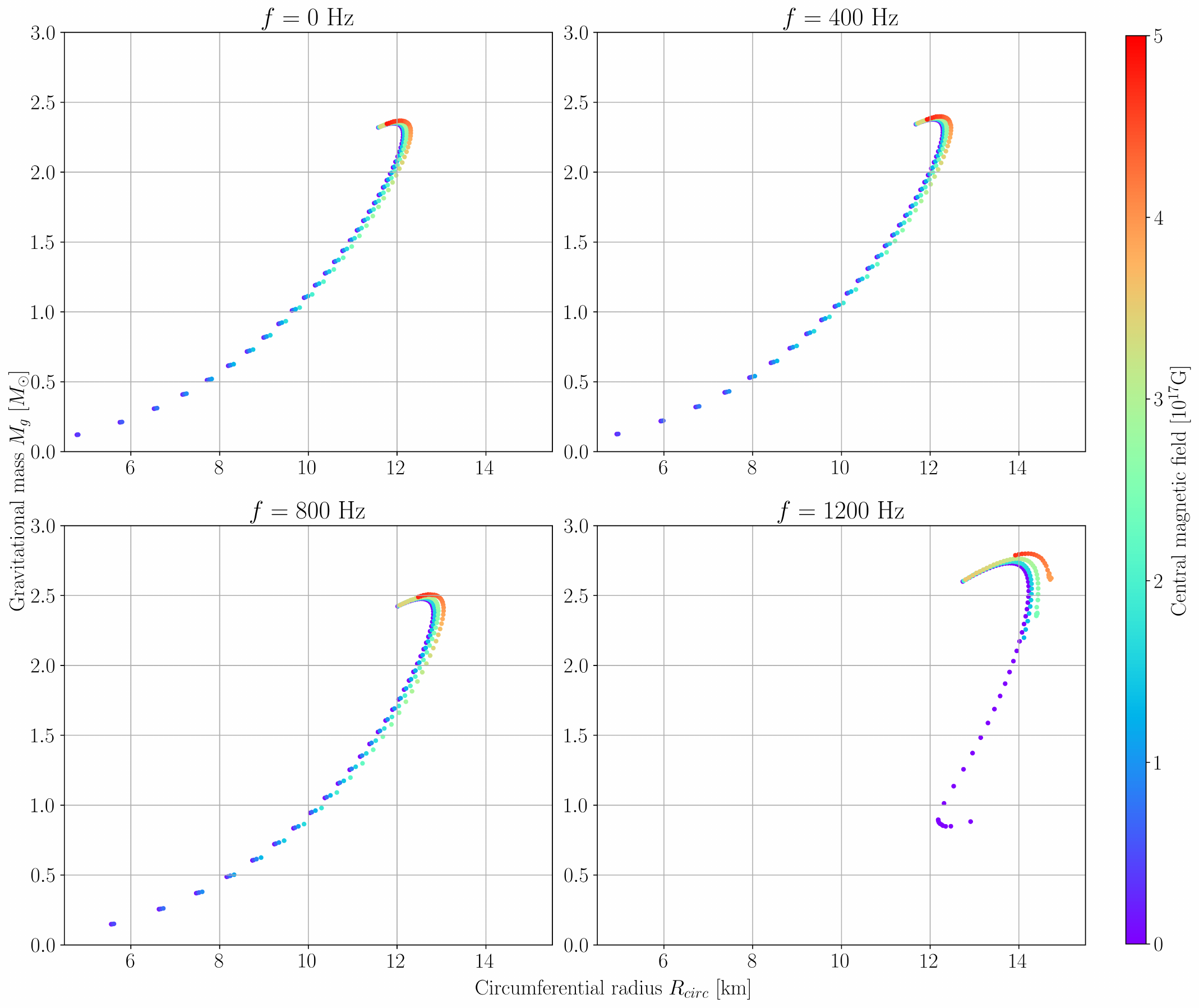}}
\caption{Gravitational mass $M_{\rm g}$ versus circumferential radius $R_{\rm circ}$ for different rotational frequencies in each panel, with the colors indicating the value of the central magnetic field, as indicated in the color bar.}\label{MR}
\end{figure*}
In Fig.~\ref{MR}, we show the gravitational mass $M_{\rm g}$ as a function of the circumferential radius $R_{\rm circ}$ for each configuration. 
In each panel of rotational frequency is shown that the low mass configurations ($M \lesssim M_\odot$) are not affected by the magnetic field. The influence of the magnetic field on gravitational mass increases gradually in the more massive configurations. The maximum gravitational mass $M^{\rm max}_{\rm g}$ and the corresponding radius increases with increasing the magnetic field in each panel. We also find that the configurations with higher rotational frequency are stable with larger $M^{\rm max}_{\rm g}$.
For relatively small masses ($M<M_\odot$) and low rotational frequencies, our model obeys the mass-radius relation $M\simeq 4\pi/3 \rho_0 R^3$, characterizing self-bound stars with density $\rho_0$ at zero pressure \citep{Haensel_1986}.  

\begin{table*}[h!]
\centering
\caption{Structural parameters of the configuration with the maximum gravitational mass in different models.}\label{table:1}
 \begin{tabular}{|c||c|c|c|c|c|c|}
  \hline
  $\textit{f}$ (Hz) & $B_{\rm c}$ ($10^{17}$ G) & $M_{\rm g} (M_\odot)$ & $M_{\rm b} (M_\odot)$ & $R_{\rm circ}$ (km) & $a$ &$|E_{\rm EB}|/A  $ (MeV)\\
  \hline\hline
   \multirow{8}{8pt}{\centering 0} & 0 & 2.35 & 2.92 & 11.92 & 1 & 184\\ 
   & 1.05 & 2.35 & 2.92 & 11.9 & 1.0 & 184\\ 
   & 2.43 & 2.36 & 2.93 & 11.97 & 1.01 & 184\\
   & 3.10 & 2.36 & 2.94 & 12.03 & 1.02 & 184\\
   & 3.84 & 2.36 & 2.94 & 12.03 & 1.03 & 184\\
   & 4.51 & 2.37 & 2.95 & 12.09 & 1.04 & 183\\
   & 5.11 & 2.37 & 2.95 & 12.21 & 1.05 & 183\\
  \hline\hline
   \multirow{2}{8pt}{\centering 400} & 0 & 2.38 & 2.96 & 12.05 & 1.03 & 184\\
   & 1.03 & 2.38 & 2.95 & 12.1 & 1.03 & 184\\
   & 1.74 & 2.38 & 2.96 & 12.1 & 1.03 & 184\\
   & 3.12 & 2.39 & 2.97 & 12.15 & 1.05 & 183\\
   & 3.78 & 2.39 & 2.98 & 12.22 & 1.06 & 183\\
   & 4.5 & 2.40 & 2.98 & 12.22 & 1.07 & 183\\
   & 5.15 & 2.40 & 2.99 & 12.34 & 1.08 & 182\\
  \hline\hline
   \multirow{2}{8pt}{\centering 800} & 0 & 2.48 & 3.07 & 12.54 & 1.11 & 182\\
   & 1.05 & 2.48 & 3.10 & 12.55 & 1.12 & 182\\
   & 2.45 & 2.49 & 3.10 & 12.59 & 1.13 & 182\\
   & 3.87 & 2.5 & 3.10 & 12.65 & 1.15 & 182\\
   & 4.6 & 2.51 & 3.11 & 12.70 & 1.16 & 182\\
   & 5.11 & 2.51 & 3.11 & 12.93 & 1.20 & 180\\
  \hline\hline
   \multirow{2}{8pt}{\centering 1200} & 0 & 2.73 & 3.38 & 13.08 & 1.36 & 179\\
   & 1.04 & 2.73 & 3.38 & 13.84 & 1.36 & 179\\
   & 2.44 & 2.75 & 3.40 & 13.90 & 1.38 & 179\\
   & 3.83 & 2.78 & 3.43 & 14.10 & 1.43 & 178\\
   & 4.51 & 2.80 & 3.46 & 14.21 & 1.46 & 177\\
   & 4.97 & 2.80 & 3.43 & 14.6 & 1.55 & 171\\
  \hline
 \end{tabular}
\end{table*}
The frame corresponding to $f= 1200\,$Hz in Fig.~\ref{MR} shows that the configurations with gravitational masses less than specific values depending on the magnetic field strength do not exist. In other words, only massive SQSs may exist as strongly magnetized, rapidly rotating objects where the binding energy is in balance with the magnetic and rotation energy of the stars. The minimum gravitational mass limit of the rapidly rotating model and Keplerian configurations of SQS are discussed in the companion paper.

\begin{figure*}
\centering
\makebox[\linewidth][c]{\includegraphics[width=368.432bp]{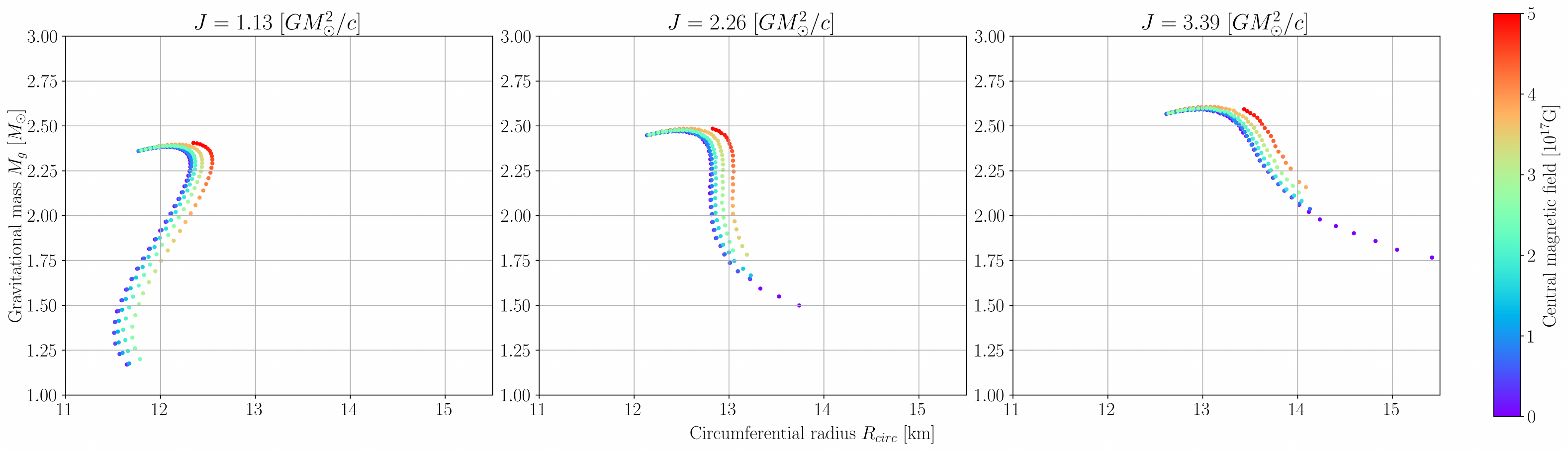}}
\caption{Gravitational mass $M_{\rm g}$ versus circumferential radius $R_{\rm circ}$ with different angular momentum ($J = 1.13, 2.26, 3.39\, [{GM_\odot^2/c}]$) in each panel with the colors indicating the value of the central magnetic field.}\label{MRJ}
\end{figure*}

We investigate the stability of our model against axisymmetric perturbations by examining the derivative of the mass with respect to the radius at constant angular momentum $J$ in which the $(dM/dR)_J$ must be greater than zero. In other words, an increase in the stellar radius at a fixed angular momentum should give increased stellar gravitational mass. This criterion indicates that the star has the ability to withstand minor deformations and oscillations without undergoing collapse or mass loss. 
To find the maximum stable configuration in each model, we examine the mass-radius relation at constant angular momentum $J$. An example is illustrated in Fig.~\ref{MRJ} which displays three frames representing the angular momentum 1.13, 2.26, and 3.30~${GM_{\odot}^2/c}$. The maximum gravitational mass in each sequence corresponds to the maximum stable configuration in each magnetic field and angular momentum.

\begin{figure*}
\centering
\makebox[\linewidth][c]{\includegraphics[width=368.432bp]{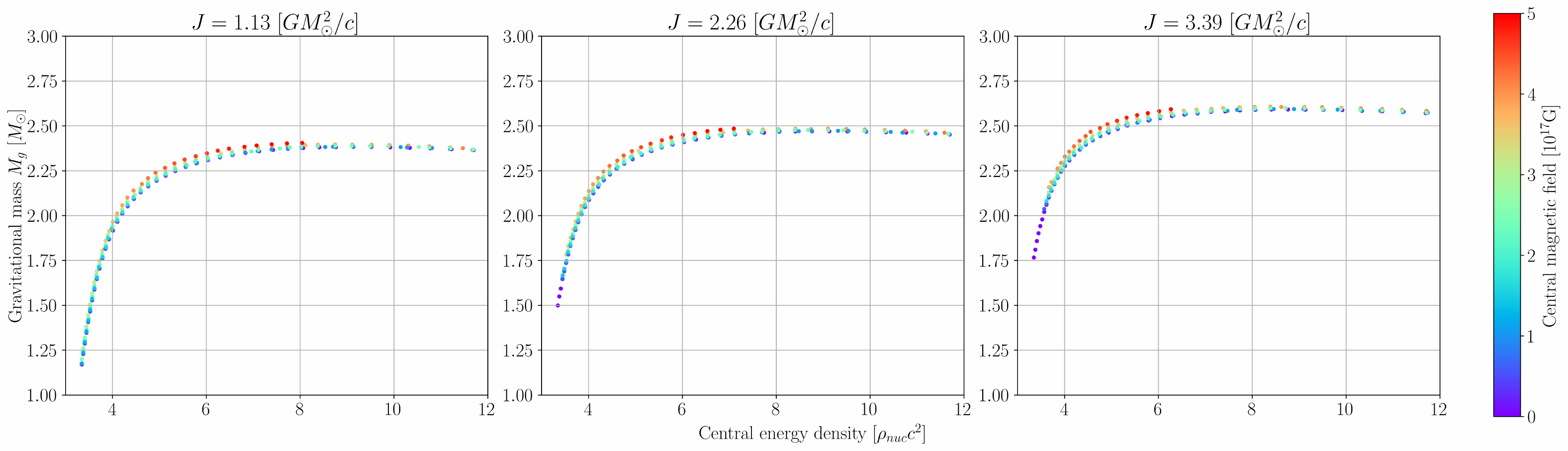}}
\caption{Gravitational mass $M_{\rm g}$ as a function of central energy density in the unit of [$\rho_{\rm nuc} c^2$] with different angular momentum ($J = 1.13, 2.26, 3.39\, [{\rm GM_\odot^2/c}]$)in each panel with the colors indicating the value of the central magnetic field.}\label{MEJ}
\end{figure*}

In Fig.~\ref{MEJ} is shown the gravitational mass as a function of central energy density for different values of magnetic field and angular momentum. For a given central energy density, the gravitational mass increases with increasing magnetic field and angular momentum. This behavior can be attributed to the increased pressure and density gradient near the center of the star, which can support a larger gravitational mass. In particular, as the magnetic field strength increases, the central energy density of the maximum gravitational mass also increases. Similarly, as the angular momentum increases, there is a shift towards higher gravitational masses at a given central energy density. In Table~\ref{table:1}, we show the numerical results of the maximum stable configurations of SQS in each model of magnetic field and rotational frequency.

\begin{figure*}
\centering
\includegraphics[width=223.880bp]{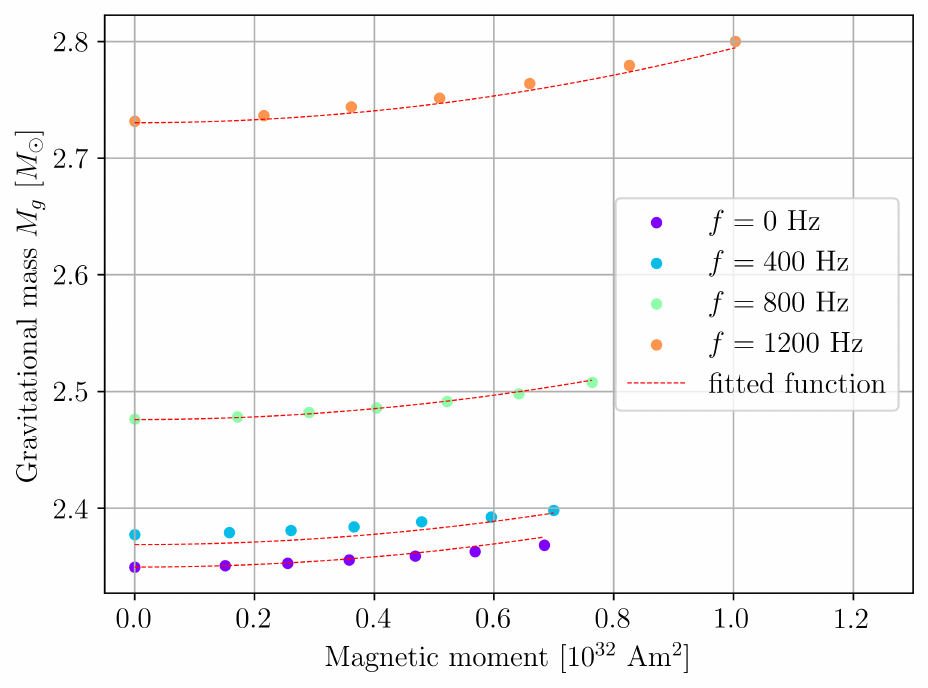}
\caption{Gravitational mass $M^{\rm max}_{\rm g}$ as a function of the magnetic moment $\mu$ at different rotational frequencies $f$. Each color represents one rotational frequency. The dotted lines show the fitted function.}\label{mBf}
\end{figure*}

In Fig.~\ref{mBf} are shown the maximum gravitational mass $M^{\rm max}_{\rm g}$ versus the corresponding magnetic moment ($\mu$) in each model. The magnetic moment is in the unit of ${\rm A\cdot m^2}$ and rotational frequency $f$ in the unit of Hz. We fit a function on the data as follows, 
\begin{equation}
    \frac{M_{\rm g}^{\rm max} (\mu, f)}{M_{\rm g}^{\rm max}(0, 0)} \simeq (1+a \mu^2)(1+b f^c), \label{eqmg}
\end{equation}
where $a$, $b$, and $c$ are the constant values. In order to increase the accuracy of our fitting, we create a sequence of rotational frequencies ranging from $0$ to $1200\,$Hz, with a step size of $200\,$Hz. We obtained coefficients of $a = 2.34\times10^{-2}\, {\rm A^{-2}m^{-4}}$, $b = 6.88 \times 10^{-10}\, {\rm s}^{2.72}$ and $c=2.72$ with the accuracy of $10^{-3}$.

We find a variation in the maximum gravitational mass $M^{\rm max}_{\rm g}$ of non-magnetized SQS from $2.35\,M_\odot$ to $2.73\,M_\odot$, as the rotational frequency rises from zero to $1200\,$Hz. In the non-rotating model $M^{\rm max}_{\rm g}$ increases from 2.35 to $2.37\,M_\odot$ by increasing the magnetic field from 0 to $5\times10^{17}\,$G. However, at each rotational frequency ($f=400$, $800$, and $1200\,$Hz), $M^{\rm max}_{\rm g}$ reaches $2.40\,M_\odot$, $2.50\,M_\odot$, and $2.80\,M_\odot$, respectively, at the strongest central magnetic field in each model.

\subsection{Comparison with observations}

The most recent detection of millisecond ``black widow'' pulsar, \emph{PSR J09520607} estimates the gravitational mass of $M_{\rm g} \simeq 2.35\, M_\odot \pm 0.17$ and the dipole surface magnetic field of $B\simeq 6 \times 10^7\,$G, with the period of 1.4~s corresponding to the rotational frequency of $\simeq 700\,$Hz \citep{Romani_2022}. We compute the maximum gravitational mass of $2.35\, M_\odot$ for the non-magnetized, non-rotating SQS and $2.38\, M_\odot$ for a non-magnetized SQS with the rotational frequency of $400\,$Hz, which is compatible with that estimated for \emph{PSR J09520607}.
Also, our model for the pulsar with the rotational frequencies of $800$ and $1200\,$Hz estimates $M^{\rm max}_{\rm g}\geq 2.5 \, M_\odot$, which can explain the recent detection of \emph{GW190814} by the LIGO/Virgo collaboration, where the gravitational mass of the lower mass object in the binary is estimated between $2.5\,M_\odot$ and $2.67 \, M_\odot$. Our magnetized SQS model can explain the Chandra X-ray detection which estimates the mass and radius of magnetar \emph{SGR J1745-2900}, $2\, M_\odot$ and $13.7\,$km, respectively. The surface magnetic field of this source is $2\times 10^{14}\,$G \citep{magnetar2015,magnetar2020}. Furthermore, there are a significant number of compact objects with a mass of about $2\,M_\odot$ which can be explained using our models, e.g. \emph{PSR J1614-2230} with a mass of $M=1.908 \pm 0.016\,M_\odot$ and \emph{PSR J0348+0432} with a mass of $M= 2.01\pm 0.04 \,M_\odot$, both detected in the pulsar-white dwarf binary systems \citep{Dem:2010:Nature:,Zhao:2015:}.

\subsection{Magnetic and rotational deformation}\label{Sdeformation}
\begin{figure*}
\centering
\includegraphics[width=354.300bp]{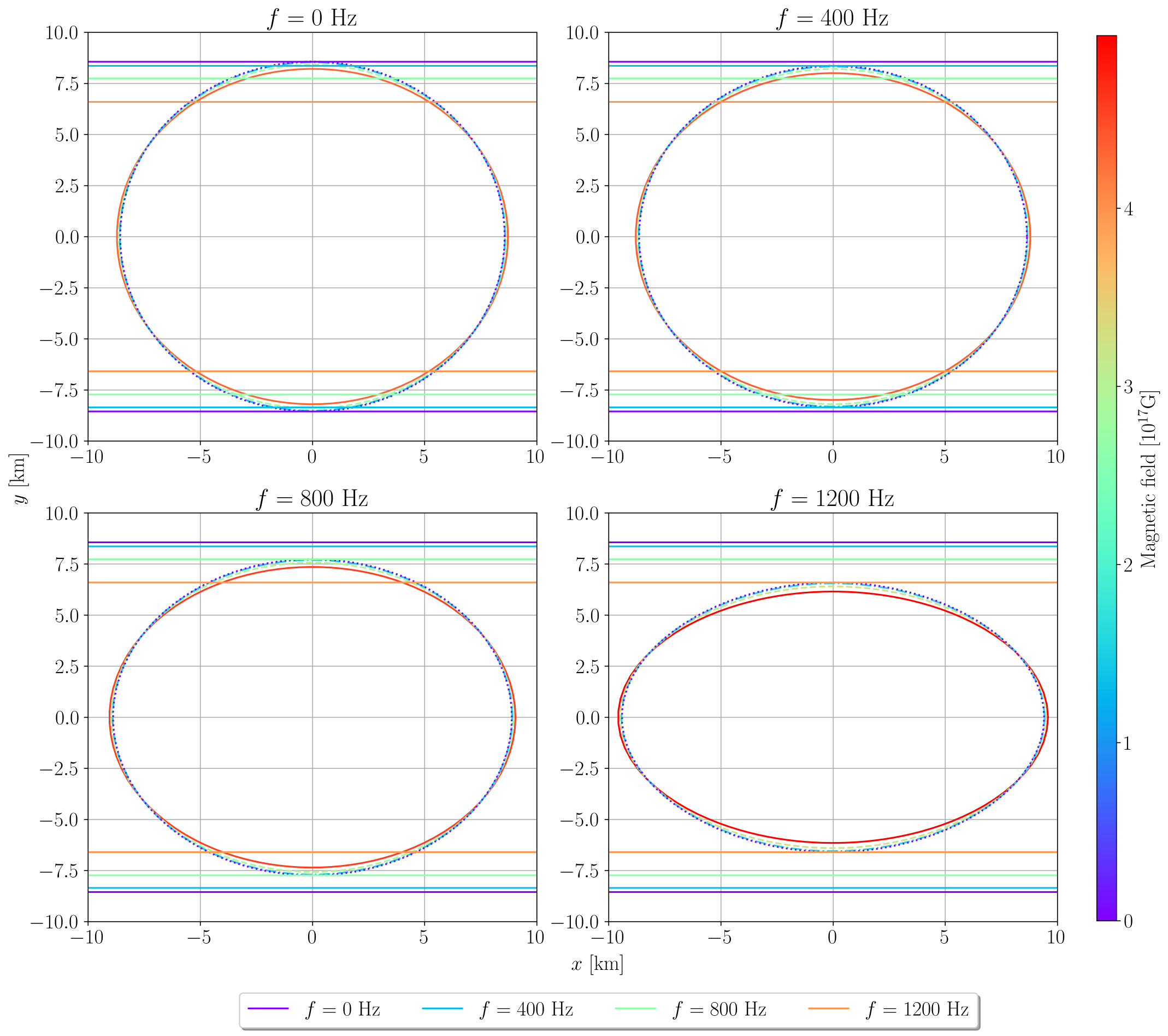}
\caption{The stellar shape of SQS in different computed models. The color of the ellipses indicates the magnitudes of the central magnetic field. Straight lines which are at the same positions in all panels indicate the polar radius of the non-magnetized configuration in each rotational frequency. 
}\label{def}
\end{figure*}
\begin{figure*}
 \centering
  \includegraphics[width=246.578bp]{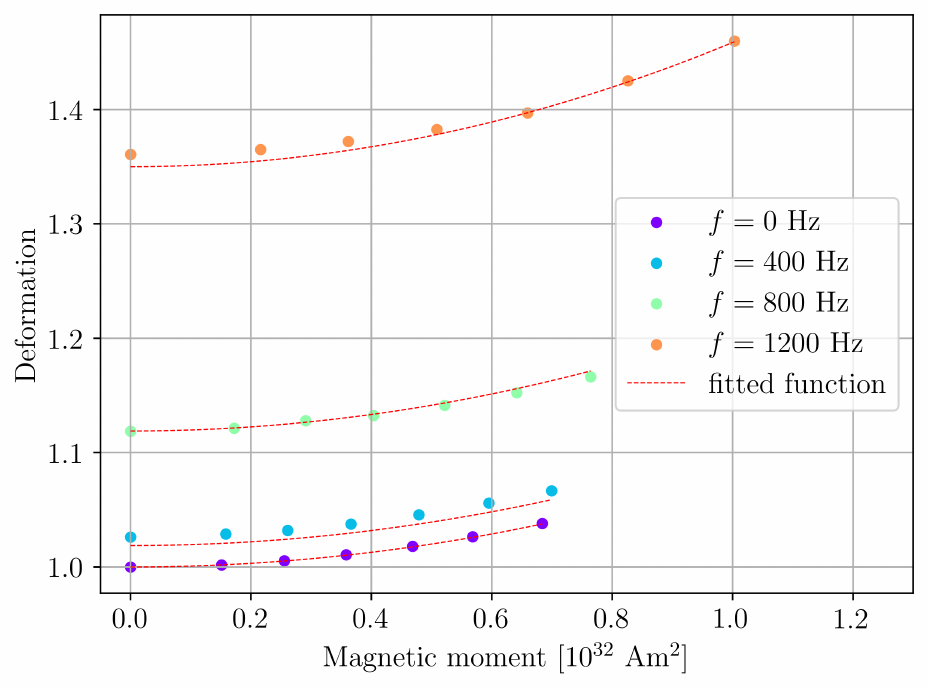}
 \caption{Deformation parameter $a$ versus magnetic moment $\mu$ in different rotational frequencies for the maximum stable configurations of SQS in each computed model.}
 \label{Deform}
\end{figure*}
Both rotation and magnetic field depending on their strength may break the spherical symmetric shape of the star and cause the deformation. The magnetic deformation depends on the magnetic field configuration. We consider the poloidal magnetic field, where radial $B_{\rm r}$ and poloidal  $B_{\rm \theta}$ components of the magnetic field are the non-vanishing components.

In Fig.~\ref{def} is shown the surface of the SQS configuration with the maximum gravitational mass in each computed model with different rotational frequency and central magnetic field. The horizontal lines, which are at the same positions in four panels with different rotational frequencies, indicate the position of the polar radius of the non-magnetized configuration, where the magenta line corresponds to the configuration with $f=0$ and the orange line corresponds to the configuration with $f = 1200\,$Hz. Following the horizontal lines, we find  that the star becomes more oblate with increasing rotational frequency, as expected. In each panel, colors indicate the magnetic field. In each rotational frequency, the polar radius decreases slightly with increasing the magnetic field. The oblateness does not change significantly with the magnetic field, but the change is significant with increasing rotational frequency.

We compute the deformation parameter $a=R_{\rm eq}/R_{\rm pol}$ as the ratio of the equatorial radius $R_{\rm eq}$ to the polar radius $R_{\rm pol}$, for the configurations with the maximum gravitational mass. In Fig.~\ref{Deform} is shown the deformation parameter $a$ as a function of the magnetic moment $\mu$ in the unit of ${\rm A \cdot m^2}$, with the different rotational frequencies. This plot quantitatively demonstrates the impact of the magnetic field and rotational frequency on the shape of the SQS. We find the deformation parameter as a function of magnetic and rotational frequency,
\begin{equation}
    \frac{a(\mu, f)}{a(0,0)} \simeq (1+\Tilde{a} \mu^2) (1+\Tilde{b} f^{\Tilde{c}})
    \label{aBf}
\end{equation}
where $\Tilde{a} = 8.03\times 10^{-2}$ ${\rm {A}^{-2}m^{-4}}$, $\Tilde{b}=2.22\times 10^{-10}$ ${\rm s}^{2.66}$, and $\Tilde{c}=2.66$. We generate a series of rotational frequencies for a more accurate fitting function, similar to Eq.~\ref{eqmg}. The accuracy of the fitting is in the order of $10^{-3}$.
 
The deformation parameter $a$ can be found in the fifth column of Table~\ref{table:1}. We find that the maximum deformation parameter $a=1.55$ corresponds to magnetized rotating SQS with $B_{\rm c} \simeq 5 \times 10^{17}\,$G and $f = 1200\,$Hz. 
\subsection{The total energy of SQS}\label{energy}
The total energy $E_{\rm tot}$ of the SQS measured at infinity is a sum of the internal, $E_{\rm int}$ and external, $E_{\rm ext}$ energies: $E_{\rm tot}=E_{\rm int}+E_{\rm ext}$. As mentioned earlier, the SQS model is considered to consist of SQM inside and a dipolar magnetic field outside of the star. The strength of the dipolar magnetic field decreases with the distance from the star.

To calculate the internal energy, we integrate the energy-momentum tensor $T^{\rm \mu \nu}$ over the volume of SQS using the LORENE library. The external magnetic energy is computed using the following method.

Given that outside of the SQS there is only a magnetic field without matter, we neglect the general relativistic impacts on the external energy. The energy density of the magnetic field can be expressed as $B^2/2\mu_0$ where $\mu_0$ is the constant vacuum magnetic permeability. Since $\boldsymbol{\nabla}\times{\boldsymbol{B}}=0$, we can introduce a magnetic potential $\phi$ such that $\boldsymbol{\nabla}{\phi}=\boldsymbol{B}$. This allows us to define the external energy of the star as follows, 
\begin{equation}
E_{\rm ext}=\int_V \frac{1}{2\mu_0}\boldsymbol{\nabla} \phi\cdot\boldsymbol{\nabla} \phi =\frac{1}{2\mu_0}\int_{dV} da \left(\boldsymbol{n}\cdot\boldsymbol{\phi}\boldsymbol{\nabla} \phi \right),
\end{equation}
 where $V$ represents the domain outside of SQS. According to the Gauss theorem,
 \begin{equation}
\boldsymbol{\nabla} \cdot{\left(\phi\boldsymbol{\nabla}{\phi}\right)}=\phi\boldsymbol{\nabla}^2{\phi}+\boldsymbol{\nabla}{\phi}\cdot\boldsymbol{\nabla}{\phi}=\boldsymbol{\nabla}{\phi}\cdot\boldsymbol{\nabla}{\phi}.
 \end{equation}
The triple integral can be simplified to a double integral over the surface of the star. By using the axisymmetric coordinates, this double integral can be further reduced to a single integral. As a result, we can consider $\phi$ as the only effective parameter, which is assumed to be produced by the magnetic dipole moment $\mu$:
\begin{equation}
    \phi(r,\theta)=\frac{\mu \cos{\theta}}{4\pi r^2}.
\end{equation}

We use $33$ points lying on the surface of the star, sampled from LORENE, and use the Lagrange interpolation to construct the function describing the stellar surface. The integral over the stellar surface is calculated using the trapezoid rule.

In Table~\ref{table:2}, we show the values of the external and total energies for non-magnetized ($B_{\rm c} = 0$) and strongly magnetized ($B_{\rm c} = 5\times 10^{17}\,$G) configurations in each rotational frequency. The magnetic field and rotational frequency of SQS affect both the maximum gravitational mass and the total energy. We find that the contribution of magnetic energy to total energy is less than $1\%$. The total energy increases about 4\% from non-rotating ($f = 0$) to fast-rotating ($f = 1200\,$Hz) model. 

In our models, the total energy of the SQS is approximately $5.5 \times 10^{47}\,$J ($\approx 10^{54}\,$ergs). For comparison, the energy of a Type II supernova is estimated up to $10^{51}\,$ergs \citep{supernovae2015}, while the energy of a quark-nova is estimated at approximately $10^{52}\,$ergs \citep{Ouyed_2015, ouyed2022}.
\begin{table*}
\centering
\large
 \caption{The total and external energies associated with the maximum gravitational mass for both non-magnetized $B = 0$ and strongly magnetized $B = 5\times10^{17}\,$G configurations in each rotational frequency.}\label{table:2}
 \begin{tabular}{|c|c|c|c|c|}
  \hline
  $f$ (Hz)&$B_{\rm c}$($10^{17}$\,{\rm G})&$M_{\rm g} (M_\odot)$&$E_{\rm ext}$ ($10^{46}\,$J)&$E_{\rm tot}$ ($10^{46}\,$J)\\
  \hline\hline
  \multirow{2}{8pt}{\centering 0}&0&2.35&0.0&54\\
  &5& 2.37&0.05&55\\
  \hline
  \multirow{2}{8pt}{\centering 400}&0&2.38&0.0&54\\
  &5& 2.40&0.05&55\\
  \hline
  \multirow{2}{8pt}{\centering 800}&0&2.48&0.0&54\\
  &5&2.51&0.02&55\\
  \hline
    \multirow{2}{13pt}{\centering 1200}&0&2.73&0.0&55\\
  &5&2.80&0.05&56\\
  \hline
 \end{tabular}
\end{table*}
\subsection{Binding energy and compactness}\label{bind}

We study the relationship between the total binding energy $E_{\rm BE}=(M_{\rm g}-M_{\rm b})c^2+E_{\rm ext}$, with $M_{\rm b}$ denoting the baryon mass of the star and the compactness parameter $\beta=M_{\rm g}/R_{\rm circ}$, where $\beta$ is in the units of $M_\odot/{\rm km}$.

We find a linear relation between total binding energy and compactness for the given EOS model,
\begin{equation}
    \frac{E_{\rm BE}}{M_{\rm g}} = - \beta.
\end{equation}
This equation is approximately applicable to all computed models of rotational frequency and magnetic field.

To evaluate the performance of our model, we compare the binding energy computed in our model with other studies. We find that the binding energy computed in our model is comparable with those computed in \cite{SQSbinding, Drago_2020, Jiang_2019}. For instance, the total binding energy of SQS in a confined density-dependent mass model (CIDDM) is $\sim0.57$ with the gravitational mass of $2.07 M_{\odot}$ gives $|E_{\rm EB}|/M_{\rm g} \simeq 0.28$ \citep[see Table~II in][]{Jiang_2019} which is slightly higher than the $\sim0.24$ computed in our model. 

In the last column of Table~\ref{table:1}, we show the total binding energy per baryon number $A$ in the range of $171 \,{\rm MeV} \leq|E_{\rm BE}|/A \leq 184\,{\rm MeV}$, which confirm the stability of our model.
\section{Summary and conclusions}\label{CONCL}
We investigated the impact of magnetic fields and rotational frequency on the structural parameters and energy of strange quark stars. We constructed a model of a non-magnetized non-rotating SQS, as well as several models of magnetized rotating strange quark star, by varying the central magnetic field and rotational frequency. In each model, we made a sequence of 51 configurations by changing the central enthalpy in the range of $0.01\, c^2$ to $0.51\, c^2$ with a spacing of $0.01\, c^2$.

The equation of state of strange quark matter is computed using the density-dependent MIT bag model, taking into account the Landau quantization effect in Fermi relations arising from the strong magnetic field interior of compact objects. 

To calculate the structural parameters, we used the $3+1$ formalism to solve the axisymmetric Einstein field equations, which resulted in four elliptic partial differential equations. The \texttt{Et\_magnetisation} class of the LORENE library was employed to solve the four elliptic partial differential equations. 

The gravitational mass of the strange quark star is computed as $2.35\, M_\odot$ in a non-magnetized non-rotating model. The gravitational mass slightly increased with the increase in the strength of the magnetic field. Additionally, the rotating models showed larger maximum gravitational masses. In the model with the rotational frequency of 1200~Hz and the central magnetic field $5\times 10^{17}\,$G, the maximum gravitational mass reached $2.80 \,M_\odot$.

The value of gravitational mass computed in our model was comparable to observational data, e.g. the predicted values by \cite{Abbott:2020:} for \emph{GW190814} with a mass between $2.5\, M_\odot$ and $2.67 \, M_\odot$ and the gravitational mass of pulsars such as \emph{PSR J0952-0607} with a mass of $2.35\, M_\odot$.

We established a relation between the maximum gravitational mass, magnetic moment, and rotational frequency by fitting a function to the computed data. The details of the fitted function were discussed in \S\ref{MgR}.

In addition, we studied the magnetic and rotational deformations of the strange quark star. The deformation parameter was defined as a ratio of the equatorial radius to the polar radius. Our model showed that the maximum deformation of the strange quark star is 1.55 in the fast-rotating strongly magnetized model with the rotational frequency of 1200~Hz and central magnetic field of $5\times 10^{17}\,$G. We found the deformation parameter as a function of magnetic moment and rotational frequency as discussed in \S\ref{Sdeformation}. 

We estimated the total energy of the strange quark star in the order of $10^{54}\,$ergs. It was also shown that the contribution of the external energy to total energy is a fraction of a percent. 

We found that the binding energy of the strange quark star is a linear function of compactness. In our proposed models, the ratio of binding energy to gravitational mass is approximately $0.24$, which aligns with other theoretical studies, different models, e.g. with the density-dependent mass model (CIDDM) studied by \cite{Jiang_2019}. Furthermore, the binding energy per baryon number of the strange quark star in our model is computed to be between $171-184\,$MeV, with compactness of approximately $0.25\, M_\odot/{\rm km}$ across all configurations.

We also showed that in the fast-rotating model with rotational frequency 1200~Hz, the strange quark star approaches the minimum gravitational mass. The study of the minimum gravitational mass and Keplerian configuration of the non-magnetized strange quark star is presented in the companion paper to this work. 

A limitation of our current model lies in accurately calculating the critical configurations of strongly magnetized, rapidly rotating stars. Future studies will focus on refining the model to enable more precise analysis and characterization of these challenging cases. Additionally, we plan to incorporate a broader range of equations of state in our investigation, with the goal of identifying global relations between the structural parameters of compact stars.

\Acknow{This project was funded by the Polish NCN (grant No. 2019/33/B/ST9/01564). 
FK is supported by the Polish NCN (Preludium-22 no. 2023/49/N/ST9/01398) 
and the program Vouchers for Universities in the Moravian-Silesian Region
 (registration number CZ.10.03.01/00/23-042/0000390), within the project Accreting Magnetized Neutron Stars. 
 M\v{C} also acknowledges the Czech Science Foundation (GA\v{C}R) grant No.21-06825X and the support 
 by the International Space Science Institute (ISSI) in Bern, which hosted the International Team project
 no. 495 (Feeding the spinning top) with its inspiring discussions. 
 We thank Prof. Wlodek Kluzniak for advice and discussions, 
 and the LORENE team for the possibility of using the code.}

\end{document}